\documentclass{optica-article}
\usepackage{comment}

\journal{opticajournal} 

\articletype{Research Article}
\usepackage{svg}
\usepackage{lineno}

\begin{document}

\title{Shaping entangled photons through thick scattering media using an advanced wave beacon}

\author{Ronen Shekel,\authormark{1,$\dag$} Ohad Lib,\authormark{1,$\dag$} and Yaron Bromberg \authormark{1,*}}

\address{\authormark{1}Racah Institute of Physics, The Hebrew University of Jerusalem, Jerusalem 91904, Israel \\
\authormark{$\dag$}These authors contributed equally to this work.}

\email{\authormark{*}yaron.bromberg@mail.huji.ac.il} 

\begin{abstract*} 
Entangled photons provide transformative new paths in the fields of communication, sensing, and computing. However, when entangled photons propagate through a complex medium such as a biological tissue or a turbulent atmosphere, their correlations are scrambled. Using wavefront shaping to compensate for the scattering and retrieve the two-photon correlations is challenging due to the low signal-to-noise ratio of the two-photon signal. While previous works partly addressed this challenge by using feedback from a strong classical laser beam that co-propagates with the entangled photons, such methods frequently depend on assumptions about the complex medium, limiting the applicability of quantum wavefront shaping. In this work, we propose and demonstrate a new feedback mechanism that is inspired by Klyshko's advanced wave picture, in which the classical laser beam counter-propagates with one of the entangled photons and co-propagates with the other. The new Klyshko feedback allows compensation of scattering in thick samples and even in situations where each photon propagates through a different scattering medium. Since the advanced wave picture applies whenever optical reciprocity is valid, such Klyshko optimization can be utilized across a wide range of configurations, offering a robust and alignment-free setup. We therefore believe this protocol will open the door for real-world applications of quantum wavefront shaping. 
\end{abstract*}

\section{Introduction}
Photonic quantum technologies, including quantum communications \cite{sidhu2021advances}, imaging \cite{bowen2023quantum}, and computing \cite{kok2007linear, bartolucci2023fusion}, use single and entangled photons to encode quantum information, typically in the form of two-dimensional quantum bits. Recently, the use of high-dimensional encoding, which has been shown to increase the information capacity and resilience to noise of quantum technologies, is gaining significant interest \cite{cerf2002security, ecker2019overcoming, erhard2020advances}. Specifically, high-dimensional spatially entangled photons that can be generated via spontaneous parametric down-conversion (SPDC) and efficiently controlled using spatial light modulators (SLMs) have been successfully used for both fundamental tests of quantum mechanics \cite{dada2011experimental, designolle2021genuine} and for advancing the performance of quantum technologies \cite{mirhosseini2015high, lib2023resource, Cameron2024Adaptive}.

One of the main challenges of practical realizations of high-dimensional quantum technologies is scattering and aberrations \cite{lib2022quantum}. When spatially encoded photons travel through a complex medium such as a turbulent atmosphere, biological tissue, or a multimode fiber (MMF) - their spatial correlations are randomized, scrambling the information they carry. Since scrambling of information limits applications in long-range quantum communication and quantum imaging through turbid media, it is critical to develop methods that can compensate for the scattering of spatially entangled photons. 

Experiments from the past decade \cite{defienne2014nonclassical, huisman2014controlling, defienne2016two, defienne2018adaptive, Peng2018manipulation, valencia2020unscrambling, lib2020pump, courme2023manipulation, devaux2023restoring, shekel2023pianoq} have shown that this scrambling can be overcome, by using methods from the field of wavefront shaping such as iterative optimization\cite{vellekoop2007focusing}, or transmission matrix reconstruction \cite{popoff2010measuring}. Remarkably, good control over the quantum signal in a complex medium even allows to program its interference \cite{pinkse2016programmable} and create general quantum circuits \cite{leedumrongwatthanakun2020programmable, goel2024inversedesign, makowski2024large}. 

A major obstacle when trying to transfer wavefront shaping methods to entangled photons is the inherently weak signals associated with quantum light, which render the feedback for scattering compensation slow and noisy. This challenge can be partly overcome by using feedback from a bright laser beam that co-propagates with the entangled photons, by adding an auxiliary laser to the setup \cite{defienne2014nonclassical, defienne2016two, defienne2018adaptive, leedumrongwatthanakun2020programmable, devaux2023restoring, courme2023manipulation, makowski2024large} or by utilizing the classical pump laser that generates the entangled photon pairs \cite{lib2020real, lib2020pump, shekel2021shaping}. However, such methods often rely on assumptions about the thickness, dispersion, or memory effect range of the complex medium, restricting the applicability of quantum wavefront shaping.

In this work, we propose and demonstrate an advanced wave approach for compensating scattering of spatially entangled photons through arbitrary scattering media. We rely on Klyshko’s advanced wave picture (AWP), which utilizes the optical reciprocity theorem to relate the quantum correlations between photons created via SPDC and the classical intensity of an appropriate, classical, advanced wave \cite{klyshko1988simple, belinskii1994two, arruda2018klyshko}. To optimize the correlations between the photons after propagating through a scattering medium, we replace one of the single photon detectors in the experiment with a classical laser, named the Klyshko beam. The Klyshko beam propagates back towards the nonlinear crystal, where instead of hitting the crystal, we reflect it back towards the setup with a mirror, and measure its intensity at the plane of the second detector. Through Klyshko’s AWP, the intensity of the advanced wave is guaranteed to follow the correlations between the entangled photons, regardless of the properties of the scattering medium. Thus, by optimizing the intensity of the Klyshko beam using an SLM, we also optimize the correlations between the entangled photons. We demonstrate the flexibility of this approach by passing each photon through different, thick diffusers and compensating for their scattering within and beyond the memory effect range and to one or two spots. By allowing for fast scattering compensation through general scattering media, our method paves the way for implementing wavefront shaping techniques for quantum technologies.  

\section{Klyshko's advanced wave picture}
\begin{figure}[ht!]
    \centering
    \includegraphics[width=\linewidth]{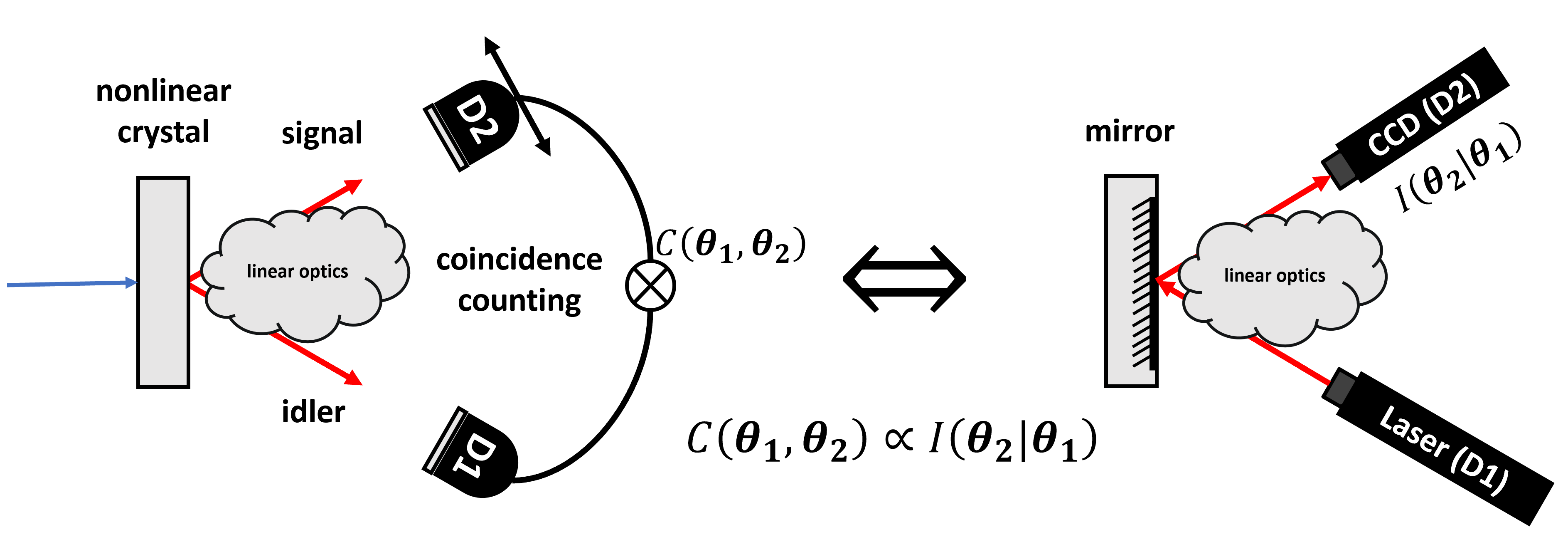} 
    \caption{\label{fig:concept}\textbf{Klyshko's advanced wave picture.} For any reciprocal linear optical setup, the average two-photon coincidence rate $C$ at angles $\theta_1$ and $\theta_2$ of photons generated via SPDC is equivalent to the average intensity $I$ of a classical advanced wave in a Klyshko configuration. In the Klyshko configuration, a classical laser originating from angle $\theta_1$ propagates backward through the optical setup, reflects off a mirror in the crystal plane, and then propagates through the optical setup again before finally being measured at angle $\theta_2$.}
\end{figure}

The advanced wave picture, developed by D. N. Klyshko \cite{klyshko1988simple, belinskii1994two, arruda2018klyshko}, is a simple approach for describing correlations between photons created via SPDC. In Klyshko's AWP the average coincidence counts $C$ between two single-photon detectors in a two-photon experiment are mapped to a classical intensity measurement, by replacing one of the detectors with a classical light source.  Light emitted from this source is directed back through the optical setup towards the nonlinear crystal and is reflected back from it towards the second detector. According to the AWP, if the classical source occupies the same optical mode as the mode probed by the detector it replaced, the intensity $I$ measured by the second detector will be proportional to the average coincidence rate measured in the actual two-photon experiment. In Fig. \ref{fig:concept} we pictorially depict these corresponding pictures for a particular configuration where one single photon detector measures a photon with a transverse angle $\theta_1$ and the second detector measures a photon with a transverse angle $\theta_2$. 

The AWP has been used to intuitively explain several experiments, such as quantum optical coherence tomography \cite{abouraddy2002quantum} and ghost imaging \cite{pittman1995optical}. It has also been used for simulating quantum state engineering \cite{zhang2014simulating} and has been demonstrated in the context of imaging \cite{aspden2014experimental}. In this work, we take full advantage of this correspondence in a scenario where the signal-to-noise ratio is critical. Using feedback from a strong classical signal set in a "Klyshko configuration", we optimize quantum SPDC correlations when propagating through a complex medium. We refer to this optimization process as \textit{Klyshko optimization}.

\section{Results} 
\subsection{Experimental setup}
\begin{figure}[ht!]
    \centering
    \includegraphics[width=\linewidth]{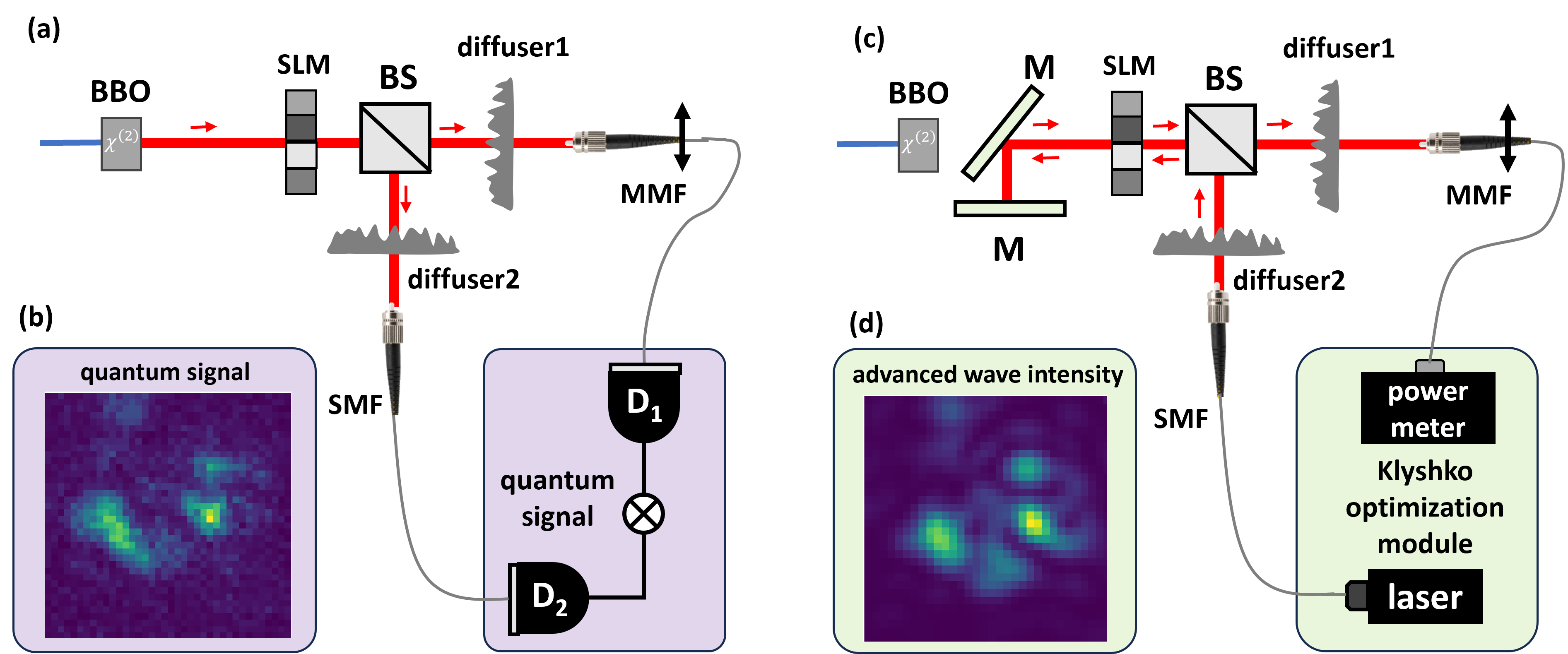} 
    \caption{\label{fig:setup}\textbf{Experimental setup.} (a) In the quantum configuration, SPDC photons are generated using a BBO crystal, pass through two different diffusers, and are collected by an MMF and an SMF placed at the far-field of the diffuser for measuring the two-photon angular correlations. An SLM is used to optimize the correlations after the propagation through the diffusers. (b) A sample two-photon speckle measured for a weak diffuser, before the optimization protocol is applied. (c) The Klyshko setup is identical to the two-photon setup, with two additional mirrors (colored green) that back-reflect the Klyshko beam back to the setup. In this setup, instead of both fibers being connected to single-photon detectors, the SMF is connected to a laser and the MMF to a power meter. (d) Intensity measurements using the Klyshko configuration with the same weak diffuser, before the optimization protocol is applied. M = mirror. BS = beam splitter, D = single photon detector.}
\end{figure}

Our experimental setup is depicted in Fig. \ref{fig:setup}a. We generate spatially entangled photons using SPDC in the thin crystal regime, forming a superposition of photon pairs propagating in opposite transverse angles \cite{walborn2010spatial}. We study the general case in which each photon passes through a different diffuser. After the photons scatter by the diffusers we collect them using optical fibers placed at the far-field of the diffusers. We measure the angular correlation between the photons by registering the rate of coincidence events, i.e. simultaneous detection of the two photons, as a function of the transverse position of one of the fibers. The two-photon angular correlation map of photons passing through the diffusers exhibits a two-photon speckle pattern \cite{peeters2010observation}, as depicted in Fig. \ref{fig:setup}b. 

The AWP points us to an analogous classical experiment that will reproduce the same result, albeit with a much higher signal-to-noise ratio. To switch the experimental setup to the Klyshko configuration, we connect the stationary collection fiber to a laser, instead of to a single photon detector. It is important for this fiber to be a single mode fiber (SMF), such that the detected mode is identical to the mode excited by the laser. Additionally, we replace the nonlinear crystal with a mirror, using a flip mount (Fig. \ref{fig:setup}c). Once the mirror is aligned with respect to the nonlinear crystal, the measurement of the laser intensity is performed with the same collection fiber that collected the SPDC photons. As seen in Fig. \ref{fig:setup}d, the classical speckle pattern measured in the classical Klyshko experiment resembles the two-photon speckle, as expected by the AWP. 

We stress the point that once the mirror is placed in the crystal plane, the alignment between the classical and quantum signal is automatically established irrespective of the optical setup and detector positions, since the very same fibers both collect the quantum signal and perform the classical experiment. Also, the alignment of the mirror is rather straightforward: When no diffuser is present, we position both fibers such that strong SPDC correlations are observed. We then move to the Klyshko configuration, and align the mirror such that the second fiber will collect a maximal amount of light. Further details regarding the experimental setup are given in the supplementary information. 

\subsection{Klyshko optimization}
Taking advantage of the correspondence between scattering of the classical Klyshko beam and the entangled photons, we use an SLM to first optimize the intensity of the classical advanced wave, enjoying its high signal-to-noise ratio. The AWP guarantees that, without any adaptation, the same optimized phase mask that focuses the classical beam will also localize the quantum correlations between the SPDC photons.

\begin{figure}[ht!]
    \centering
    \includegraphics[width=\linewidth]{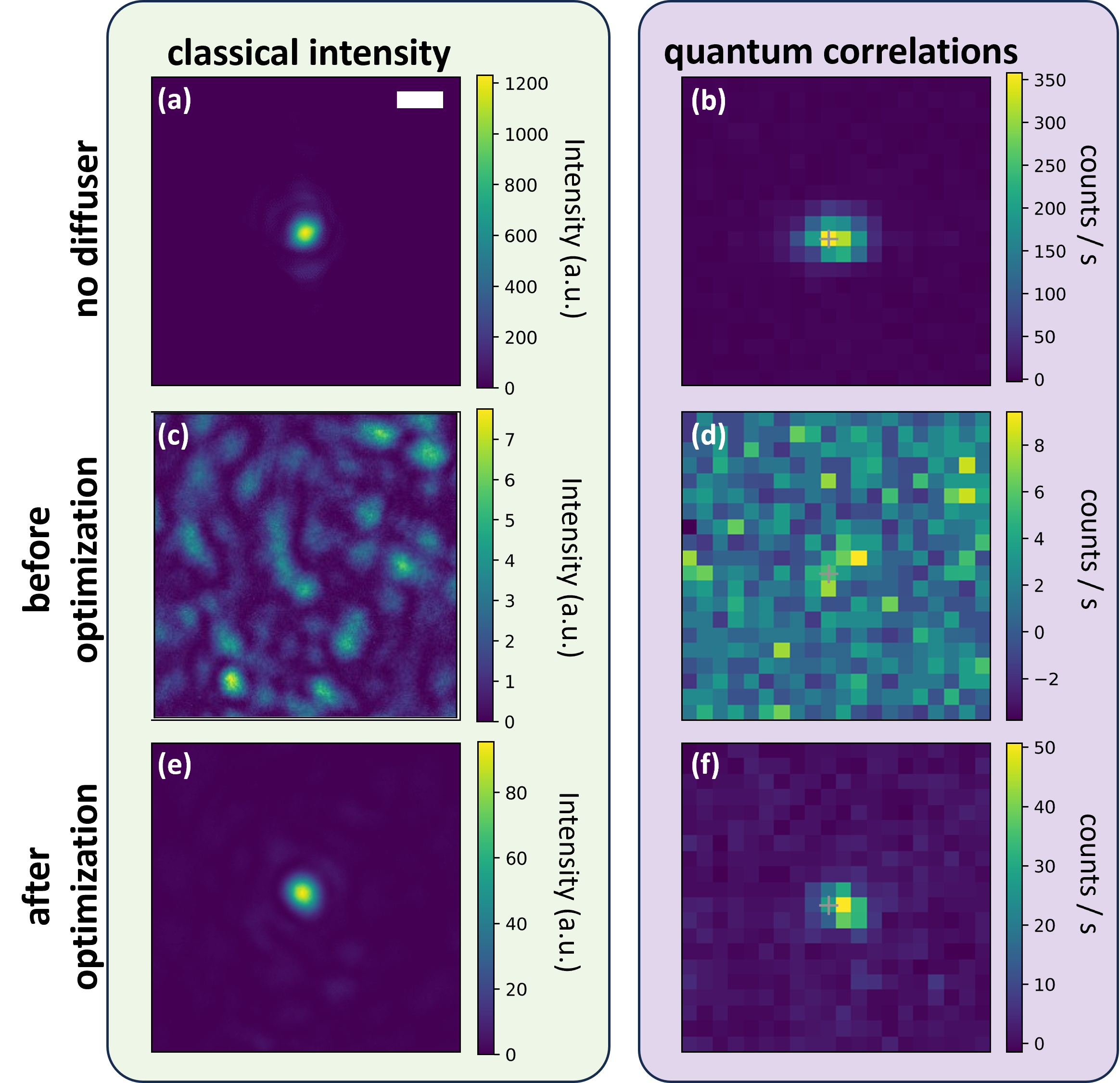} 
    \caption{\label{fig:optimization}\textbf{The optimization process.} When no diffuser is present, the laser is focused (a) and the two-photon correlations are localized (b). After we place the diffusers, the laser and entangled photons are scattered and show classical (c) and two-photon (d) speckle patterns, respectively. After applying a classical optimization protocol to the bright laser beam, it is re-focused (e). The SPDC correlations are simultaneously re-localized for the same SLM pattern found in the classical optimization (f). All panels have the same scale, where the white scale bar in (a) represents 150 $\mu m$, equivalent to $1.5$ mrad. The gray plus signs in panels b,d,f represent the location of the collection MMF during the optimization process.}
\end{figure}

We begin by measuring both the intensity of the advanced wave and the two-photon correlations when no diffuser is present, as depicted in Fig. \ref{fig:optimization}a, b. Then, we add two different diffusers, each in the path of one of the entangled photons. As depicted in Fig. \ref{fig:optimization}c, d, both the intensity of the classical Klyshko beam and the quantum correlations of the entangled photons are strongly scattered. We use the SLM to optimize the advanced wave, enhancing its intensity at a chosen focal spot (Fig. \ref{fig:optimization}e). We finally return to the SPDC configuration and measure the correlations between the photons, observing a clear enhancement of the coincidence peak (Fig. \ref{fig:optimization}f). The simultaneous scattering compensation of both signals is due to the correspondence between the classical advanced wave, which is scattered by the first diffuser, reflected at the crystal plane, and then scattered again by the second diffuser, and the two entangled photons that are scattered from one diffuser each. Further details regarding the optimization process are provided in the supplementary information. 

We note that while for the optimization process we collect the classical signal using the same fiber collecting the SPDC signal, the classical pictures in Fig. \ref{fig:optimization}a,c,e are obtained using a camera positioned in the same plane as the collection fiber. The fact that we do not need the camera for the optimization process could potentially allow for a significant speedup of the optimization process, using a photodiode and a high-speed SLM \cite{tzang2019wavefront}. 

We also note that the correspondence between the advanced wave and SPDC signals is not perfect. For instance, the classical signal has a narrower focal spot and is scattered more significantly than the SPDC signal. We discuss these deviations from the AWP correspondence in the supplementary information.

Thanks to optical reciprocity, the mathematical correspondence between the intensity of the classical Klyshko beam and the quantum correlations of the entangled photons, enables our optimization approach to work efficiently, regardless of the configuration of the scattering medium. As shown above, even though each photon passes through a different diffuser, a scenario that is relevant for entanglement distribution \cite{yin2017satellite}, the optimization is straightforward. Performing the same task using a regular beacon laser would be more complicated and involve either separate optimization runs for each diffuser, or a more complex cost function to optimize both diffusers simultaneously. In addition, in contrast to the pump shaping method, this method works also when using a thick diffuser. Indeed, in the results shown in Fig. \ref{fig:optimization}, one of the diffusers is a thick diffuser, as we quantify below by studying its memory effect.

\subsubsection{Off-axis optimization and the memory effect}
\begin{figure}[ht!]
    \centering
    \includegraphics[width=\linewidth]{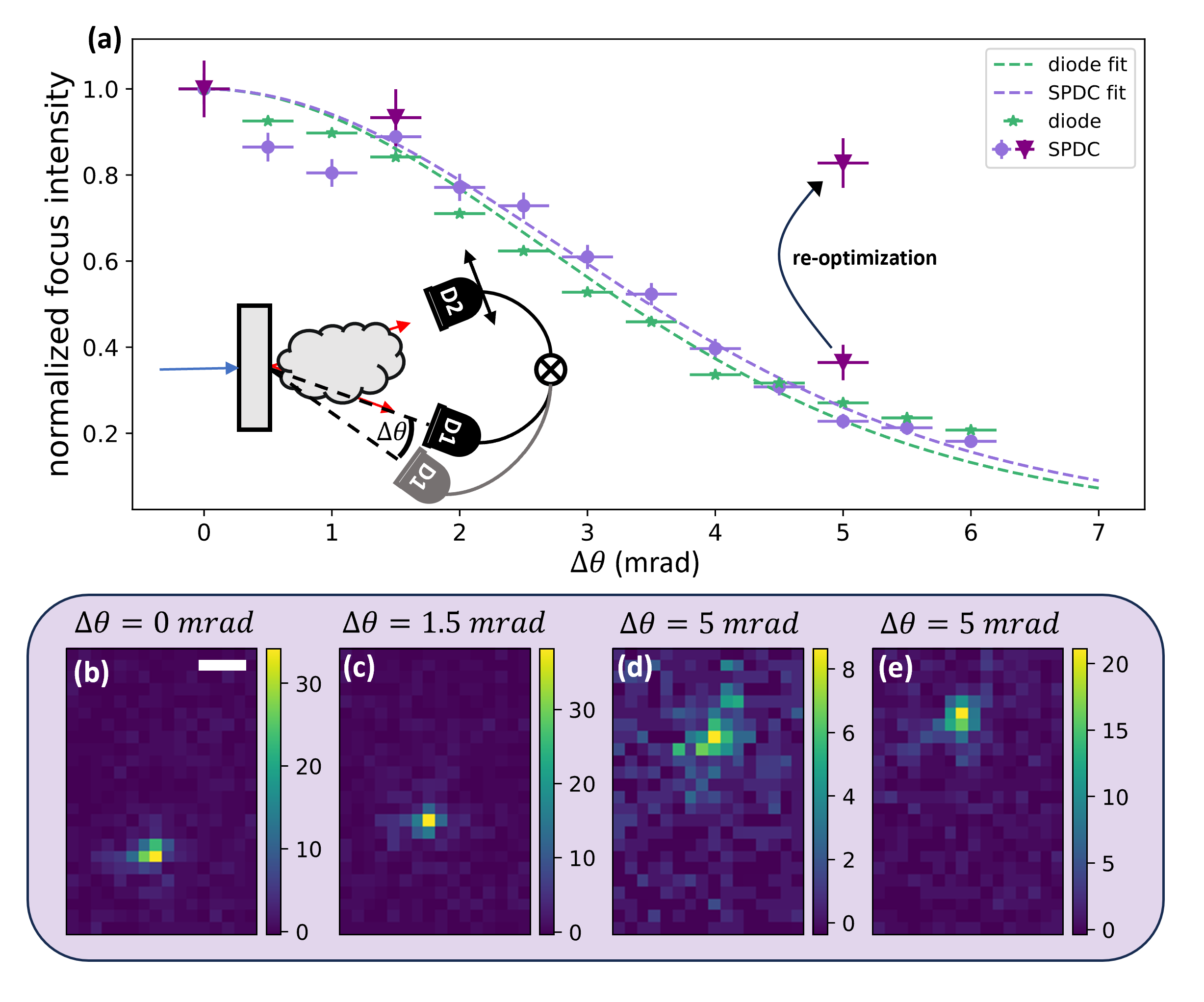} 
    \caption{\label{fig:memory}\textbf{Memory measurements and off-axis optimization.} (a) We quantify the memory effect using the Klyshko beam and the entangled photons, for the diffuser configuration used in Fig. \ref{fig:optimization}. Both measurements yield similar memory ranges. (b) Optimized two-photon angular correlations, measured when the static detector is positioned on the optical axis. (c) Two-photon angular correlation when the static detector is slightly shifted by $\Delta\theta = 1.5mrad$, within the memory range, such that the correlations remain localized. (d) Correlation measurements after the static detector is further shifted, now by $\Delta\theta = 5mrad$, outside of the memory range. The angular correlations are no longer localized, and accordingly, the correlation peak is greatly reduced. (e) The strong correlations are revived by applying the Klyshko optimization again, this time in the new position of the static detector ($\Delta\theta = 5mrad$). The purple triangles in (a) quantify the process visualized in (b)-(e). Panels (b-e) have the same scale, where the white scale bar in (b) represents $2$ mrad. The units of the color bars are counts / s.}
\end{figure}

The classical optical memory effect refers to the phenomenon in which a small change $\Delta\theta$ in the illumination angle onto a scattering sample results in an identical speckle pattern at the far-field of the sample, shifted by  $\Delta\theta$ \cite{feng1988correlations, freund1988memory}. It is often used for characterizing scattering samples \cite{schott2015characterization} and imaging through scattering layers \cite{bertolotti2012non, katz2014non}. In wavefront shaping applications, in particular, the memory effect allows scanning the optimized focal spot using the same SLM correction pattern that was found in the optimization process \cite{bertolotti2022imaging}. The angular range of the memory effect can be quantified by focusing the light to a spot, scanning $\Delta\theta$, and measuring the intensity ratio of the shifted spot to the original one \cite{katz2012looking}. Similarly, when focusing two-photon correlations through a scattering medium, when one detector is fixed and the other detector scans the transverse far-field plane, one observes a focused spot. If the fixed detector is now translated by some $\Delta\theta$ smaller than the angular range of the memory effect (as depicted in the inset in Fig. \ref{fig:memory}a), the focus will shift, this time by $-\Delta\theta$ due to the angular anti-correlations between the photons\cite{peeters2010observation,defienne2018adaptive}.

To quantify the angular range over which the angular two-photon correlation can stay localized using the same SLM pattern, we measured the angular memory range with the entangled photons and with the classical Klyshko beam, for the diffuser configuration used in Fig. \ref{fig:optimization}. The experimental results presented in Fig. \ref{fig:memory}a are fitted to a theoretical model \cite{feng1988correlations} of $\frac{I\left(\Delta\theta\right)}{I\left(0\right)}=\left[\frac{\Delta\theta/\theta_{0}}{\sinh\left(\Delta\theta/\theta_{0}\right)}\right]^{2}$, and yields very similar results of $\theta_0=2.2\pm0.4$ for the classical Klyshko beam and $2.3\pm0.4$ for the entangled photons signal. This is also expected in configurations where both photons propagate through the same thick diffuser, since the propagation of the Klyshko beam towards the crystal and from the crystal is equivalent to the two-photon propagation. In contrast, when the classical laser co-propagates with the entangled photons as in \cite{defienne2018adaptive}, the classical laser beam exhibits a broader memory range.  

To highlight the versatility and ease of alignment of our Klyshko optimization, we optimize the two-photon angular correlations of the entangled photons when the detectors are placed off the optical axis, outside the memory effect range. To optimize angular correlations outside of the memory range using the conventional beacon laser method that co-propagates with the entangled photons, one would need to realign the beacon beam to the different desired angles. In contrast, using our method, optimizing for these different angles does not require any additional alignment, since the Klyshko beam is automatically aligned with the desired angle of detection. 

In Fig. \ref{fig:memory}b-e we demonstrate deterioration of the focus and the re-optimization, in a separate realization. We begin by showing optimized two-photon angular correlations (Fig. \ref{fig:memory}b). Next, we shift the angular position of the static detector by $\Delta\theta=1.5mrad$, within the memory range. The two-photon angular correlations shift accordingly, and stay localized (Fig. \ref{fig:memory}c). We then further move the fixed detector ($\Delta\theta=5mrad$), and the correlations shift accordingly, yet start to delocalize (Fig. \ref{fig:memory}d). 

In order to re-optimize the correlations, we simply run again the Klyshko optimization, this time at the new positions of the fibers. The angular correlations of the entangled photons are indeed re-optimized, as depicted in Fig. \ref{fig:memory}e. The purple triangles in Fig. \ref{fig:memory}a quantify these measurements using the normalized coincidence counts before and after the re-optimization. We note that in this measurement, the normalized coincidence counts before the re-optimization are slightly higher than the normalized coincidence counts obtained in the measurement of the angular range of the memory effect, since even for the same diffuser, the memory range slightly changes for different disorder realizations. 

\subsubsection{Optimizing a complex cost function}
Optimizing quantum light with a complex cost function is difficult since several pixels must be measured simultaneously, which degrades the signal-to-noise ratio. Since Klyshko optimization uses classical feedback, this can be achieved by using a simple camera. Instead of maximizing the intensity entering the collection fiber, we use a flip mirror to send the advanced wave to a camera positioned at the same plane. We could now perform a more complex optimization, just like in classical wavefront shaping. 

\begin{figure}[ht!]
    \centering
    \includegraphics[width=0.7\linewidth]{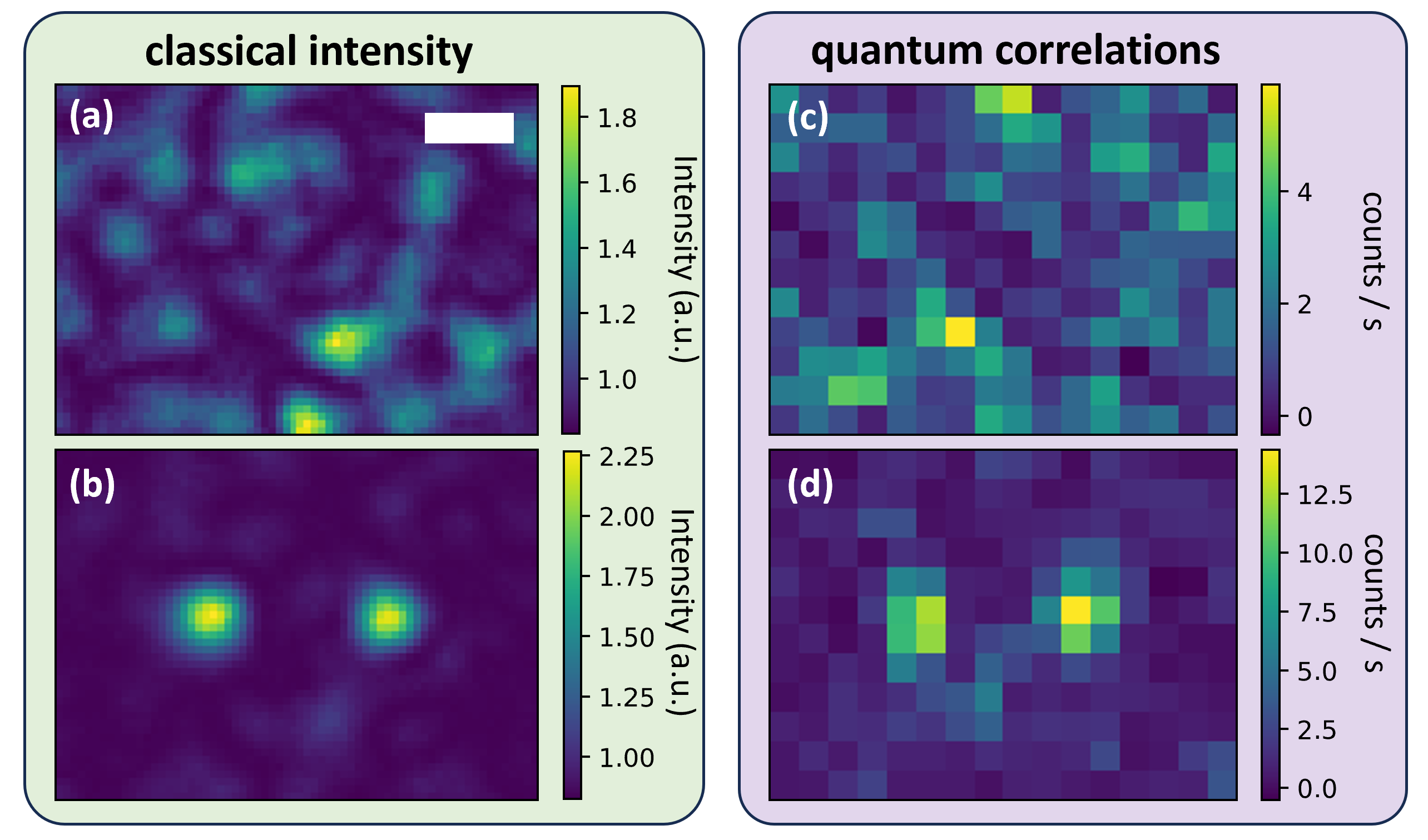} 
    \caption{\label{fig:two_Spots}\textbf{Optimizing a complex cost function.} Using a camera for feedback, we can perform optimization for more complex cost functions, such as optimizing the intensity in two spots. Performing the optimization in the Klyshko configuration, the speckle pattern in (a) is optimized to focus on two spots (b). This automatically localizes the two-photon angular correlations (c) onto two optimized spots (d). All panels have the same scale, with the scale bar in (a) representing $1.5$ mrad.}
\end{figure}

To demonstrate this, we use a camera for feedback in the Klyshko configuration, and optimize the SLM phases to get strong intensities in two different spots. Indeed, as the classical speckle pattern (Fig. \ref{fig:two_Spots}a) is optimized to two spots (Fig. \ref{fig:two_Spots}b), also the two-photon speckle (Fig. \ref{fig:two_Spots}c) is localized and shows strong correlation at the two corresponding spots (Fig. \ref{fig:two_Spots}d). 

\section{Discussion and conclusion}
In this work, we have proposed and demonstrated a method for compensating the scattering of spatially entangled photons using classical feedback based on Klyshko's AWP. By utilizing the correspondence between the scattering of the classical Klyshko beam and the entangled photons, we were able to go beyond the capabilities of recent works relying on the memory effect: we observed efficient compensating of scattering of photon pairs that propagate through different diffusers and outside the memory effect range.

While Klyshko’s AWP ensures the robustness of our method to the properties of the scattering medium, it does set some constraints on the experimental implementation. For instance, the correspondence between the effect of the nonlinear crystal and that of the mirror that back-reflects the Klyshko beam, assumes a plane-wave pump beam and the so-called thin-crystal regime, which is associated with a wide angular emission range of the SPDC pairs. In our experiment, we ensure working in this regime by using a thin $2mm$ nonlinear crystal and adding an effective pinhole on the SLM plane to take into account the finite width of our pump beam. Nevertheless, these limitations could be completely lifted. For example, by adding appropriate amplitude masks at the crystal and far-field planes, which would facilitate arbitrary pump beam shapes and crystal configurations. We further discuss this point in the supplementary information. 

As our method allows for fast classical characterization of the scattering of entangled photons, it could be readily extended beyond scattering compensation experiments by using a single-mode fiber bundle \cite{israel2017quantum} and a camera. This would allow for fast and accurate characterization of multimode random interference in complex media and efficient realizations of tailored quantum circuits on entangled photons.

Finally, thanks to the robustness and low experimental overhead of our method, we believe our technique could be used when sending multimode quantum light through multimode fibers, opening the door for high dimensional quantum communications using space division multiplexing, and for endoscopic-based quantum imaging applications.

\begin{backmatter}
\bmsection{Funding}
This research was supported by the Zuckerman STEM Leadership Program, the  Israel Science Foundation (grant No. 2497/21), and the State of Lower Saxony, Hannover, Germany. R.S. acknowledges the support of the HUJI center for nanoscience and nanotechnology, and of the ministry of innovation, science and technology. O.L. acknowledges the support of the Clore Scholars Programme of the Clore Israel Foundation.


\bmsection{Disclosures}
The authors declare no conflicts of interest.

\bmsection{Data Availability}
Data underlying the results presented in this paper are not publicly available at this time but may be obtained from the authors upon reasonable request.

\bmsection{Supplementary information}
Supplementary information is provided in a separate document, where we provide a detailed experimental setup, further details regarding the optimization process, and a detailed discussion about the deviations of the advanced wave from the quantum signal in our setup. 

\end{backmatter}

\bibliography{Main}

\end{document}


\title{Shaping entangled photons through thick scattering media using an advanced wave beacon - supplementary information}

\author{Ronen Shekel,\authormark{1,$\dag$} Ohad Lib,\authormark{1,$\dag$} and Yaron Bromberg \authormark{1,*}}

\address{\authormark{1}Racah Institute of Physics, The Hebrew University of Jerusalem, Jerusalem 91904, Israel \\
\authormark{$\dag$}These authors contributed equally to this work.}

\email{\authormark{*}yaron.bromberg@mail.huji.ac.il} 



\makeatletter
\renewcommand \thesection{S\@arabic\c@section}
\renewcommand\thetable{S\@arabic\c@table}
\renewcommand \thefigure{S\@arabic\c@figure}
\makeatother
\setcounter{figure}{0}
\setcounter{section}{0}

\section{Detailed experimental setup}
\begin{figure}[ht!]
    \centering
    \includegraphics[width=\linewidth]{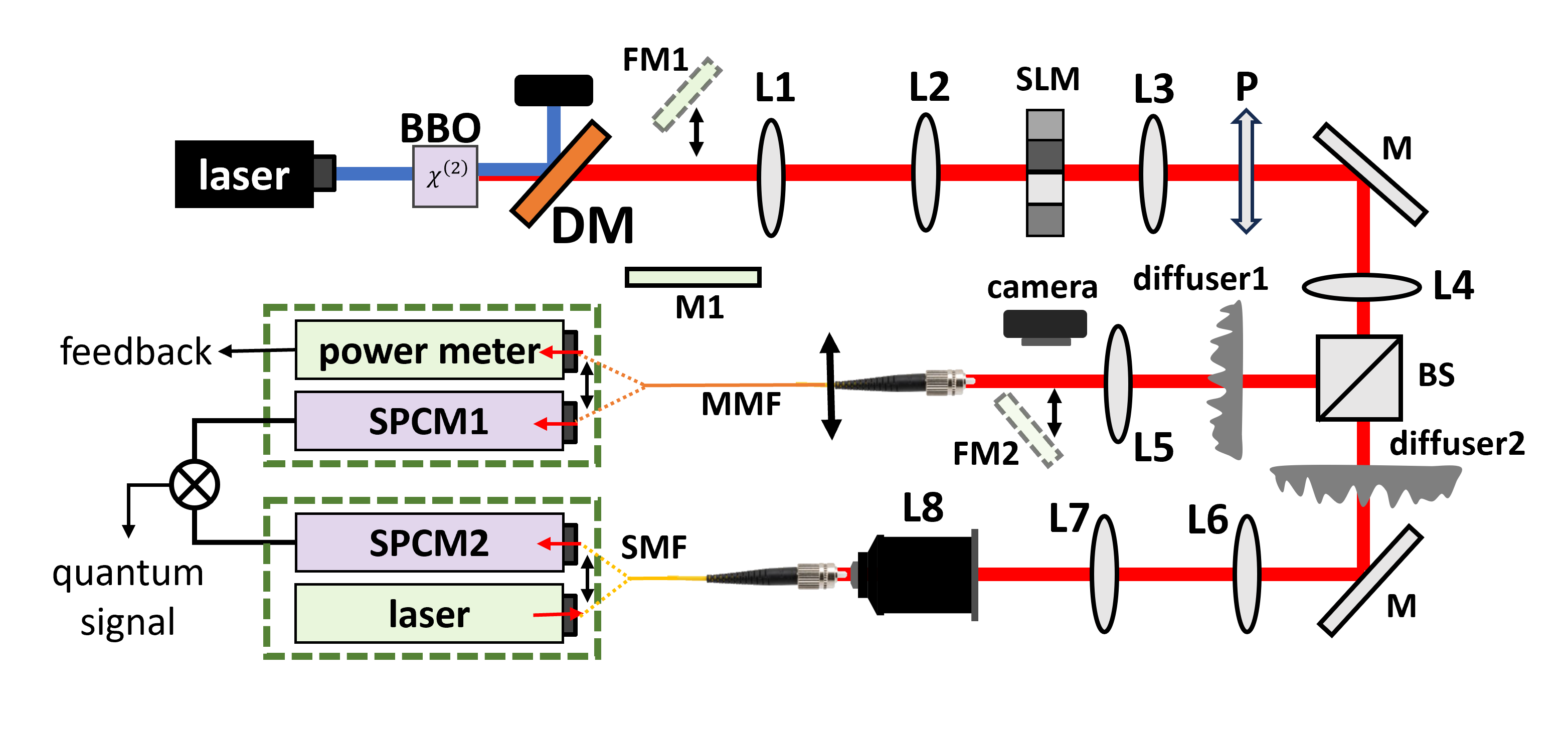} 
    \caption{\label{fig:setup_detailed}\textbf{Detailed experimental setup.} A detailed sketch of the experimental setup. DM=dichroic mirror, M=mirror, FM=flip mirror, L=lens, P=polarizer, BS=beam splitter, SPCM=single photon counting module. Elements depicted in light green (purple) correspond to elements that are introduced in the classical Klyshko (quantum two-photon) setup.}
\end{figure}

A detailed experimental setup is depicted in Fig. \ref{fig:setup_detailed}. A 2mm long BBO crystal is pumped by a continuous-wave laser (Cobolt, 125mW, $\lambda$ = 405nm). The pump profile in the crystal plane is approximately Gaussian with a waist of $w_0\approx 500\mu m$. Pairs of horizontally polarized photons are generated by the type-I SPDC process. The pump beam is separated from the entangled photons using a dichroic mirror (Semrock FF510-Di02) and sent to a beam dump. The crystal is imaged using two lenses with focal lengths $L_1=100$mm and $L_2=200$mm on the SLM (Hamamatsu LCOS-SLM X10468 series), and then imaged on the plane of the diffusers using two lenses with focal lengths $L_3=300$mm and $L_4=100$mm. The light is passed through a polarizer to discard unwanted vertically polarized light that the SLM does not modulate. On the SLM, we add a virtual pinhole, as discussed in Section \ref{section:supp_deviations} below.

The photons are probabilistically separated using a beam splitter and pass through two separate diffusers. One photon passes through diffuser1, which is an effective thick diffuser, comprised of two diffusers with a divergence angle of $0.25^\circ$ each, separated by 4.5cm (RPC photonics EDS-0.25 and EDC-0.25). This photon is then directed to the MMF (D=$50\mu m$ Thorlabs FG050LGA), placed at the far-field plane of the diffuser, obtained at the back focal plane of a lens with a focal length of $L_5=100$mm. During the correlation measurements, the MMF scans the transverse plane using two motors (Thorlabs K-cube). The second photon travels through diffuser2 which has a divergence angle of $0.5^\circ$ (Newport 10DKIT-C1). To match the core size of the SMF, it is demagnified by a factor of $\approx10X$ using lenses with focal lengths $L_6=100$mm and $L_7=200$mm, and an objective lens (Olympus 10X), before reaching the SMF (D=$5\mu m$ Thorlabs P1-780A-FC-1) in the far-field plane.

When measuring the two-photon angular correlation of the entangled photons, the two fibers are connected to single-photon detectors (Excelitas SPCM-AQRH). The coincidence counts are measured using a time tagger (Swabian, Time Tagger Ultra), with a coincidence window of $2$ns. All coincidence counts in this paper are shown after subtracting accidental counts. When performing the optimization using the Klyshko configuration, the SMF is connected to a diode laser (Thorlabs S1FC808), and the MMF is connected to a power meter (Thorlabs PM101 and Thorlabs S150C). In this configuration, the flip mirror FM1 is inserted and sends the laser to the mirror M1, which is placed in the crystal plane. When capturing images of the laser, flip mirror FM2 is inserted as well and sends the light passing through both diffusers to a camera (ZWO ASI2600).

In Fig. 2b, d in the main text, to show the resemblance between the two-photon speckle and the classical speckle, we removed diffuser1, and for diffuser2 used a weak diffuser, with a divergence angle of 0.25$^\circ$. 

\subsection{Alignment of the mirror in the crystal plane}
The only alignment needed in order to perform Klyshko optimization is the placement of a mirror in the crystal plane. This includes placing it in the correct axial plane, and setting its transverse angles. 

To align the mirror's angles, we begin in the quantum configuration, remove the diffusers, and set the transverse position of the fibers collecting the photons to the detectors such that we obtain maximal coincidence counts. We then switch to the Klyshko configuration and scan the angles of the mirror to maximize the intensity measured by the power meter. 

To align the mirror and crystal planes, we first make sure the crystal is imaged on the SLM. We do this in the quantum configuration, where we put a camera in the approximate image plane of both the crystal and SLM, and imprint a $\pi$-step on the SLM. We then find the axial positions of the SLM and camera such that both the $\pi$-step and the edge of the crystal are seen sharply. 

We then align the SLM and the mirror planes in the Klyshko configuration, by placing a camera in the plane of diffuser1, and setting a $\pi$-step on the SLM. In this configuration, the light impinges the SLM twice. We find the axial position of the mirror such that we will see a single sharp $\pi$-step. If the mirror is misaligned, we will see two $\pi$-steps, or an overlap of a sharp and a blurred $\pi$-step. 

\subsection{Future improvements to the setup}
The main advantage of using a strong classical source as feedback for optimizing a quantum signal is the large signal-to-noise ratio it yeilds, allowing fast optimization rates. While in our current setup the classical optimization rate is relatively slow, taking tens of minutes, it could easily be sped up using faster SLMs or digital micro-mirror devices (DMDs). In addition, currently, the changing of the fibers between the SPDC and Klyshko settings is done manually but could be improved by using an optical switch or an optical circulator. The usage of a flip mirror for the exchange between the quantum and Klyshko configurations could also be avoided by adding a beam sampler after the crystal, only slightly reducing the SPDC signal, as the classical signal will be strong enough also after the large reduction of intensity. Alternatively, one could use the specular reflection of the laser from the crystal facet, as demonstrated in \cite{aspden2014experimental}.

An upcoming technology in the context of quantum wavefront shaping is single photon avalanche diode (SPAD) arrays. In our context, using the classical signal, we could use a regular multi-pixel camera instead. In addition, to quickly optimize many optical modes, the SMF could be exchanged with a fiber bundle, similar to \cite{israel2017quantum}, emitting the classical light into the system from a different mode each time. 

\section{The optimization process}
When working in the Klyshko configuration, we perform the optimization using a continuous sequential protocol \cite{vellekoop2008phase} on the SLM. The SLM is divided into 363 segments, and for each segment, we find the best phase to enhance the intensity measured with the power meter. When optimizing two spots using a camera, we optimize the intensity sum in both spots, with a linear penalty such that if the spot intensities differ by $x$ percent, the cost will be reduced by $\alpha x$ percent, with $\alpha=0.2$.

We define the efficiency of the optimization as the ratio between the intensity of the optimized spot and the intensity of the spot with no diffuser. We similarly define the enhancement as the ratio between the intensity in the optimized spot and the average intensity in the speckle pattern before the optimization. 

The classical signal, which is scattered more strongly than the SPDC signal, achieves an enhancement factor of 68 and an efficiency of 7\%, while the SPDC signal achieves an enhancement factor of 20 and an efficiency of 14\%. 

We discuss the deviations between the SPDC signal and the advanced wave in Section \ref{section:supp_deviations} below. 

\section{Technical limiting factors of the Klyshko optimization} \label{section:supp_deviations}
Our method for compensating scattering of spatially entangled photons is based on Klyshko's AWP. While the AWP promises a mathematical correspondence between the classical and quantum signals, it relies on several technical assumptions. 

First, the exchange of the nonlinear crystal with a mirror assumes infinite phase-matching, i.e. that the SPDC flux is independent of the emission angles. This becomes especially important when optimizing signals far off the optical axis, or when using strong diffusers that mix many different angles. For perfect correspondence between the classical and quantum signal, one would need to add an amplitude mask in the Fourier plane of the crystal, which emulates the $sinc$ function of the phase matching, whose width depends mainly on the pump wavelength and the length of the nonlinear crystal \cite{walborn2010spatial}. 

Second, the exchange with a mirror assumes an infinite plane-wave-like pump beam. For perfect correspondence between the classical and quantum signal, one should add an amplitude mask at the image plane of the crystal, which emulates the profile of the pump. In our experiments, the mode field diameter of the SMF of the static detector is narrower than the two-photon correlation width. The classical Klyshko beam that impinges on the diffuser is, therefore, wider than the diffuser area that is covered by the entangled photons, and is scattered more strongly. For the same reason, also the focal spot obtained by the Klyshko beam is narrower than the two-photon angular correlations. Adding an amplitude mask which emulates the profile of the pump would yield the same spot size and scattering strengths for the classical Klyshko beam and the two-photon angular correlations. 

We implement this to some extent by creating a virtual pinhole using the SLM, which is imaged onto the crystal. We add opposite linear tilts inside and out of the area where the SPDC signal impinges. This attempts to discard both the specular reflection from the SLM, and also light from the diode that hits the crystal plane in areas where the pump beam did not generate SPDC photons. 

Third, to take polarization rotations into account, a Klyshko configuration for type-I SPDC must add a polarizer in the crystal plane, reflecting the fact that the SPDC photons are generated with a specific polarization. Indeed, in our setup we add a polarizer between the crystal and the diffusers, so polarization rotations that may occur when propagating through the diffusers could be ignored.

Finally, the AWP ignores spectral effects. While the diode we used for the optimization has a bandwidth of $\approx1$ nm, the SPDC bandwidth is significantly wider. In our demonstration, spectral effects were negligible. However, when working far off the optical axis, different wavelengths act differently due to geometric effects \cite{small2012spatiotemporal}. Also, when working with multimode fibers, it may prove to be challenging due to the finite spectral correlation width of multimode fibers \cite{Redding:12, Pikalek:19}. For the classical signal to perfectly emulate the quantum one, one could, in principle, use a broadband diode with the same spectrum as the SPDC signal, and some nonlinear mirror flipping the frequency of the diode symmetrically across the central SPDC frequency. More realistically, one could use narrow filters on the SPDC signal, such that the spectral dependence is negligible. 

\bibliography{Main}